\documentstyle[aps,prb]{revtex}
\begin{document}
\newcommand{\be}{\begin{equation}}
\newcommand{\ee}{\end{equation}}
\newcommand{\ba}{\begin{eqnarray}}
\newcommand{\ea}{\end{eqnarray}}
\title{The effect of Hund's Coupling on one-dimensional Luttinger Liquids} 
\author{H. C. Lee}
\address{Department of Physics, Korea University,  Seoul, Korea
  }
\author{S.-R. Eric Yang}
\address{
Department of Physics, Korea University, Seoul 136-701, Korea \\
Asia Pacific Center for Theoretical Physics, Seoul, Korea }
\draft
\maketitle 
\begin{abstract}
\indent Two one-dimensional Luttinger liquid systems coupled by Hund's coupling are studied by
the renormalization group  and the non-abelian bosonization methods.
It is found that  the Hund's coupling is always relevant 
irrespective of the repulsive interaction between electrons within each chain.
 The properties of the resulting  strong coupling fixed point are discussed.
\end{abstract}
\pacs{{\rm PACS numbers}: \hspace{.05in} 71.10.Fd, 71.10.Hf, 71.10.Pm } 
\section{Introduction}
One-dimensional (1D) electron systems provide examples where the electron correlation
effects play a dominant role.
The metallic phase of 1D interacting electron system is characterized by 
the separated  gapless charge and spin excitations, and it is called the
Luttinger liquid phase. 
Recently there has been great interest on the two coupled Luttinger liquid systems
as the model for  ladder systems,  zigzag system,  
stripe phase in cuprate 
superconductor, and  nano tubes \cite{dagotto,hamada}. 
Theoretically, the relevancy of inter-chain  hopping $t_\perp$\cite{anderson},
the opening of spin gap, and the superconducting instabilities
have been studied. \cite{strong,rice,balents,nagaosa}

In this paper, we investigate two 1D Luttinger liquid systems coupled by
Hund's coupling {\em without} interchain hopping $t_\perp$.  
This coupled system is  interesting as a first step towards 
2D ${\rm Sr}_2 {\rm Ru} {\rm O}_4$ \cite{anderson,baskaran}. 
A Hubbard chain away from half-filling
can be modelled as a Luttinger liquid for {\em arbitary} values of on-site repulsion 
$U$\cite{schulz}, and two coupled Hubbard chains in the weak coupling regime
$U, J_H < t$  can be investigated using 
the bosonization technique ($t$ is a band width, $J_H$
is Hund's coupling).
Nagaosa and Oshikawa \cite{nagaosa} investigated two coupled Luttinger liquids using a  semiclassical
analysis under the condition $ \delta \max(t, t_\perp) < |J_H|$, where $\delta$ is the
density measured from half-filling.  They showed that a spin gap opens up in this system and
that the charge fluctuations between chains are massive. 
For generic values of $t, t_\perp,
J_H$, their result is valid only near half-filling.

In present paper we investigate this model at {\em arbitrary} filling factors away from
half filling.  
We focus on the influence of Hund's coupling on the charge and 
spin excitations and its interplay with the repulsive Hubbard 
interaction.  
In order to preserve SU(2) invariance in renormalization group analysis we
use non-Abelian bosonization. 
The weak coupling phase diagram can be mapped out from the analysis of
the resulting perturbative R.G. equations.
We find that 
Hund's coupling is relevant even for infinitesimally small initial value
of $J_H$, and the system flows into a new strong coupling 
fixed point. At the strong coupling fixed point,  {\em all} spin excitations are
gapped.  As for the charge exciations, 
the symmetric charge excitations $\phi_{\rho +}, \theta_{\rho +}$ 
(See Eq.(\ref{modes}))remain gapless, while the asymmetric charge 
excitations $\phi_{\rho -}, \theta_{\rho -}$ are gapped. 
Thus, our result extends the results by Nagaosa and
Oshikawa to general filling for the case $t_\perp=0$.
Furthermore, we have investigated the dependence of the physical properties on the value
of the Hubbard repulsion,
which is not addressed by Nagaosa and Oshikawa:
We find that, even the system is in gapped phase for all $0 \le U < t$ and $ 0< J_H < t$, 
the gap (i.e. the scale where the renormalized coupling constants diverge or
become the order of the bare cut-off scale) 
strongly depends on the relative magnitude of $U, J_H$.
If  $U > J_H$ , the strong coupling
fixed point can be reached faster than the opposite case $J_H > U$.

This paper is organized as follows:
In section II, we specify the model and re-cast it in a bosonized form.
In section III, the renormalization group equations are derived using 
the operator product expansion method.
We present the renormalization group flows in section IV and conclude
with discussions in section V.
\section{Model}
We consider two Hubbard chains coupled by Hund's coupling.
\be
\label{H1}
H=-t \sum_{ij,l}  c^\dag_{l i \alpha} c_{l j \alpha}+ U
\sum_{i,l} n_{il \uparrow} n_{il \downarrow}-
\sum_{i} J_H {\bf S}_{1 i} \cdot {\bf S}_{2 i}=
H_0+H_U+H_H,
\ee
where $\alpha$ is a spin index and $l=1,2$ is a band index. $J_H$ is positive. 
Note that there is no inter-chain hopping. 
We assume the implicit normal ordering of all interaction terms in the  
Hamiltonian. For simplicity,   the band dependences of 
hopping parameter $t_{ij,l}$ and Hubbard interaction $U_l$ are ignored, two coupled chains
are considered instead of three,
and only the nearest neighbor
hopping is considered.

We investigate only  the weak coupling regime $U< t,\;\;J_H < t$, in which
 the linearization
of non-interacting electron spectrum near Fermi points and 
the perturbative treatment of interactions are legitimate.
A generic filling away from half-filling is considered,
 so that the umklapp process can be neglected.
Because the inter-chain hopping term is absent, the non-interacting energy
spectrums
of two bands  are degenerate. 
The linearization gives the decomposition 
$c_{i \alpha l}/\sqrt{a} \sim \psi_{R \alpha l}(x)\,e^{i k_F x}+ 
\psi_{L \alpha l}(x)\,e^{-i k_F x}$. $a$ is a lattice spacing. Substituting the
decomposition into $H_0$ of Eq.(\ref{H1}), we get  
\be
H_{0}=\sum_{\alpha l}\,\int d x  v_F \Big( 
\psi^\dag_{R l \alpha}\,i \partial_x \psi_{R l \alpha}-
\psi^\dag_{L l \alpha}\,i \partial_x \psi_{L l \alpha}\Big),
\ee
where $v_F=2 t a \sin (k_F a)$.
To express the Hamiltonian in the bosonized form,
it is convenient to introduce the (chiral) charge and spin current operators.
\be
J_{R,l}=\sum_\alpha \psi^\dag_{R,l,\alpha} \psi_{R,l,\alpha},\;\;
{\bf J}_{R,l}=\sum_{\alpha \beta} \psi^\dag_{R,l,\alpha} \frac{\bbox{\sigma}_{
\alpha \beta}}{2} \psi_{R,l,\beta}.
\ee
The left moving currents are defined analogously.
The operator product expansion allows us to express $H_0$
 in terms of currents \cite{boso1}.
\be
H_0=\sum_l \int dx \Big[ \frac{\pi v_F}{2} (J^2_{Ll}+J^2_{Rl})+ 
\frac{2\pi v_F}{3} ({\bf J}_{Ll}\cdot{\bf J}_{Ll}+{\bf J}_{Rl}\cdot{\bf J}_{Rl})\Big].
\ee
The Hubbard part $H_U$ can be expressed  in terms of currents alone,
 while the Hund's coupling
term $H_H$ cannot be, since 
\be
\label{spin}
{\bf S}_l/a={\bf J}_{Ll}+{\bf J}_{Rl}+e^{-2 i k_F x}\,\psi^\dag_{R l \alpha}
\frac{\bbox{\sigma}_{\alpha \beta}}{2}\,\psi_{L l \beta}+
e^{+2 i k_F x}\,\psi^\dag_{L l \alpha}
\frac{\bbox{\sigma}_{\alpha \beta}}{2}\,\psi_{Rl \beta}.
\ee
At this point, we apply Abelian and Non-abelian bosonization to 
the charge and spin part, respectively.
The phase field  $\phi_{\rho l}$ and its conjugate momentum 
$\Pi_{\rho l}$ are introduced for the charge currents \cite{boso1}.
\be
\label{para1}
J_{Rl}+J_{Ll}=\sqrt{\frac{2}{\pi}}\,\partial_x\phi_{\rho l},\;\;
J_{Rl}-J_{Ll}=-\sqrt{\frac{2}{\pi}}\,\Pi_{\rho l}.
\ee
And for the spin currents SU(2) matrix field $g_l(x,\tau)$ is introduced
\cite{witten,kz}.
\be
\label{para2}
J^a_{L l}=-\frac{i}{\pi}{\rm tr}(\partial_z g_l g_l^{-1} \frac{\sigma^a}{2}),\;\;
J^a_{R l}=+\frac{i}{\pi} {\rm tr}(g_l^{-1} \partial_{\bar{z}} g_l \frac{\sigma^a}{2}),
\ee
with $z=x+i v_F \tau$, $\bar{z}=x-i v_F \tau$.
The elements of the matrix field $g_l$ can also be expressed in terms of 
scalar field through the vertex operator construction \cite{fz}
\be
\label{vertex}
g_{\alpha \beta} \sim \pmatrix{ e^{i \sqrt{2\pi} \phi_{ \sigma l}} &
                                e^{i \sqrt{2 \pi} \theta_{\sigma l}} \cr
                               e^{-i\sqrt{2\pi} \theta_{\sigma l}} &
                               e^{-i \sqrt{2 \pi} \phi_{\sigma l}} \cr}.
\ee 
$\phi_{\sigma l}$ is the phase field of the Abelian bosonized spin currents,
and $\theta_{\sigma l}$ is the conjugate field. The above Abelian bosonized
representation  is useful in interpreting the results of R.G. 
equations physically.
For later convenience, we define the symmetric and asymmetric charge and spin modes.
\be
\label{modes}
\phi_{\rho \pm}=\frac{1}{\sqrt{2}}\Big(\phi_{\rho 1} \pm \phi_{\rho 2} \Big),\;\;
\phi_{\sigma \pm}=\frac{1}{\sqrt{2}}\Big(\phi_{\sigma 1} \pm \phi_{\sigma 2} \Big).
\ee
According to the Non-abelian bosonization rule \cite{boso1,witten,kz}
the fermion bilinear can be represented as 
\be
\label{bilinear}
\sum_{\alpha \beta}\psi^\dag_{R l \alpha}
\frac{\bbox{\sigma}_{\alpha \beta}}{2}\,\psi_{L l \beta} \sim 
{\rm tr}\big( g_l^\dag(x) \frac{\bbox{\sigma}}{2} \big)\,
e^{-i \sqrt{2\pi}\,\phi_{\rho l}(x)}.
\ee
As a result of the  parametrizations of currents and fermion bilinear
Eq.(\ref{para1},\ref{para2},\ref{bilinear}),
it is possible to express the total Hamiltonian in terms of the scalar fields
$\phi_{\rho l}$ and the SU(2) matrix fields $g_l$. 
For the renormalization group analysis,
the Euclidean Lagrangian formulation  is more convenient.
\ba
S&=&\frac{ v_c}{ 2 K_c}\,\sum_l \int d x d \tau\,\Big[
\left(\frac{\partial \phi_{\rho l}}{\partial v_c \tau} \right)^2+
\left(\frac{\partial \phi_{\rho l}}{\partial x} \right)^2 \Big]+
\sum_l S[g_l]_{WZW} \label{s0} \\
&-& \lambda_1 \sum_l \int  d x d \tau  {\bf J}_{Rl}\cdot {\bf J}_{Ll} \nonumber \\
&-&\int dx d \tau\Big[
\lambda_2 \big({\bf J}_{L1} \cdot {\bf J}_{L2}+{\bf J}_{R1} \cdot {\bf J}_{R2}
\big) +
\lambda_3 \big({\bf J}_{L1} \cdot {\bf J}_{R2}+{\bf J}_{R1} \cdot {\bf J}_{L2}
\big)\Big] \nonumber \\
&-&\lambda_4\int d x d \tau\Big[ {\rm tr}\big( g_1^\dag(x) \frac{\bbox{\sigma}}{2} \big)\,
e^{-i \sqrt{2\pi}\,\phi_{\rho 1}(x)}\,
{\rm tr}\big( g_2(x) \frac{\bbox{\sigma}}{2} \big)\,
e^{+i \sqrt{2\pi}\,\phi_{\rho 2}(x)}
+{\rm h.c} \Big]\label{perturbation}   \\
&-&\lambda_5\int d x d\tau\Big[ {\rm tr}\big( g_1^\dag(x) \big)\,
e^{-i \sqrt{2\pi}\,\phi_{\rho 1}(x)}\,
{\rm tr}\big( g_2(x) \big)\,
e^{+i \sqrt{2\pi}\,\phi_{\rho 2}(x)}
+{\rm h.c} \Big]\nonumber  \\
&+&\frac{ v_c}{ 2 K_{2}}\,\sum_l \int d x d \tau\,2\Big[
\left(\frac{\partial \phi_{\rho 1}}{\partial v_c \tau}
\frac{\partial \phi_{\rho 2}}{\partial v_c \tau} \right)+
\left(\frac{\partial \phi_{\rho 1}}{\partial x}
\frac{\partial \phi_{\rho 2}}{\partial x} \right) \Big]\label{generated} 
\ea
where
\ba
v_c=v_F\,\sqrt{1+\frac{Ua}{\pi v_F}}
\ea
and
\ba
K_c=1/\sqrt{1+\frac{Ua}{\pi v_F}}.
\ea
$S[g]_{WZW}$ is the so-called Wess-Zumino-Witten action, and we don't need its explicit
form. We just note that the Fermi velocity for the matrix field $g$ is renormalized 
by the interaction 
$v_F \to v_s=v_F (1-\frac{Ua}{2\pi v_F})$.
The initial value (bare value) of $\lambda_1$ is $2 U a > 0$, and 
the initial value of $\lambda_2$, $\lambda_3$ and $\lambda_4$ is $J_H a > 0$.
Although the initial values of $\lambda_2$, $\lambda_3$ and $\lambda_4$  
are identical they will scale differently under the R.G. flow.
The $1/K_2$ and $\lambda_5$ terms in the above action
 is absent in the bare action (the initial values are 0),  but they are 
 generated by the second order perturbation.

The action  Eq.(\ref{s0}) defines a critical fixed point of two independent
$U(1)$ Gaussian model and  two independent SU(2) $k=1$ Wess-Zumino-Witten (WZW) 
theory. The terms in Eq.(\ref{perturbation})($\lambda_i, i=1,2,3,4, 5$)
 are the perturbations to the
fixed point. We study how the perturbations influence the fixed point defined by
Eq.(\ref{s0}) via the renormalization group method in the next section.

\section{Renormalization Group Equations}
It is well-known that the renormalization group (R.G.) equations 
for the fixed point Hamiltonian perturbed by a number of scaling operators,
up to the second order in couplings,
 are determined by the scaling dimensions of the scaling operators and
 the operator product expansion (OPE)
coefficents of the scaling operators \cite{cardy}. Explicitly,
\be
\label{rg}
\frac{ d \lambda_i}{d \ \ln L}=(d - x_i) \lambda_i-\sum_{jk} 
c_{ijk} \lambda_j \lambda_k+\cdots,
\ee
where $\lambda_i$ is the coupling constants of the  scaling operator
$O_i$,
$d=1+1=2$ is the space-time dimension, and $x_i$ is the scaling dimension of the
 scaling operator.  $L$ is a cut-off length or time scale.
The OPE coefficents $c_{ijk}$ is determined by
\be
O_i(r) O_j(0) \sim \sum_k  \frac{c_{ijk}}{r^{x_i+x_j-x_k}} O_k(0).
\ee
Among the perturbations in Eq.(\ref{perturbation}), only $\lambda_4$
has a scaling dimension different from 2, which is $x_4=1+K_c$.

The relevant operator product expansions for our case are
 \cite{kz,kadanoff,fujimoto}
\ba
\label{gaussian}
e^{i \alpha \phi_{\rho l}(x,\tau) } e^{- i \alpha \phi_{\rho l}(0,0)}&\sim&
\frac{1}{|z_c|^{\alpha^2 K_c/2\pi}}\,\Big[1+ i \alpha (z_c \partial_{z_c}
 \phi_{\rho l}(0,0)+\bar{z}_c \partial_{\bar{z}_c} \phi_{\rho l}(0,0))
 \nonumber \\
&-&\alpha^2 |z_c|^2\,\partial_{z_c} \phi_{\rho l}(0,0) 
\bar{z}_c \partial_{\bar{z}_c} \phi_{\rho l}(0,0) -
\frac{\alpha^2}{2} (z_c^2 (\partial_{z_c} \phi_{\rho l})^2+
\bar{z}_c^2 (\partial_{\bar{z}_c} \phi_{\rho l})^2)+\cdots \Big],
\ea
with $z_c=x+i v_c \tau$, $\bar{z}_c=x-i v_c \tau$.

The OPE of WZW model ($k=1$)  are given by
\ba
J_{Ll}^a(z) J_{Lm}^b(w)&\sim &\frac{1}{(2\pi)^2}\,\frac{k}{2}
\frac{ \delta_{lm} \delta_{ab}}{(z_s-w_s)^2}+
\frac{1}{2\pi}\,\frac{ i \epsilon^{abc}}{z_s-w_s}\,J_{Ll}^c(w)+\cdots \label{ope1} \\
J^a_{Ll}(z)(x,\tau)\,g_m(0,0)&\sim&  \frac{ \delta_{lm} }{2\pi z_s}\,t^a  g_m
 +\cdots
\label{ope2} \\
{\rm tr}(g_l(x,\tau) \sigma^a)\,{\rm tr}(g^\dag_m(0,0) \sigma^b) &\sim&
\frac{\delta_{lm}}{|z_s|}\Big[\delta_{ab} + i \epsilon^{abc} (z_s 
J^c_{L l}(0,0)+\bar{z}_s J^c_{R l}(0,0)) \nonumber \\
&+&z_s^2 \sum_c J^c_{L l}(0,0) J^c_{L l}(0,0)+\bar{z}_s^2 
 \sum_c J^c_{R l}(0,0) J^c_{R l}(0,0) \Big]+ \cdots\label{ope3},
\ea
where $z_s=x+i v_s \tau,\;\;\bar{z}_s=x-i v_s \tau$, and $t^a$ are 
the generators of SU(2) algebra in the spin basis
$t^a=\frac{\sigma^a}{2},\;\;{\rm tr}(t^a t^b)=\frac{\delta^{ab}}{2},\;\;
[t^a,t^b]=i\epsilon^{abc}\,t^c $.
The terms of dimension higher than 2  are omitted since they
are irrelevant in our computation.
The OPE Eqs.(\ref{ope1},\ref{ope2}) are the fundamental consequencs of the 
chiral gauge invariance of WZW model. The OPE Eq.(\ref{ope3}) follows from
the fusion rule of Kac-Moody primary fields \cite{fz}.
 For level 1 $(k=1)$ SU(2) WZW theory, the product of
spin 1/2 primary fields can only give rise to the identity field and its 
descedants \cite{kz,fz}. $J$ and $J^2$ terms in Eq.(\ref{ope3}) are
the descendant fields of the identity field.  
 
Using  Eqs.(\ref{rg},\ref{gaussian},\ref{ope1},\ref{ope2},\ref{ope3}),
the R.G. equations can be computed straighforwardly. 
 The results are :
\be
\label{K}
\frac{d \frac{1}{K_c}}{d \ln L}=\frac{ \lambda_4^2}{\pi v_c
v_s}\,f_1(\frac{v_c}{v_s}),\;\;
f_1(u)=\int_0^\pi\frac{ d \theta}{(\cos^2 \theta+u^2 \sin^2 \theta)^{K_c-1}}.
\ee
\be
\label{K2}
\frac{d \frac{1}{K_2}}{d \ln L}=-\frac{\lambda_4^2}{\pi v_c
v_s}f_1(\frac{v_c}{v_s}),
\ee
\be
\label{1}
\frac{d  \lambda_1}{d \ln L}=-\frac{\lambda_1^2}{2 \pi v_s}.
\ee
\be
\label{2}
\frac{d  \lambda_2}{d \ln L}=0.
\ee
\be
\label{3}
\frac{d  \lambda_3}{d \ln L}=-\frac{1}{2\pi}\,\left( \frac{\lambda_4^2}{v_c}+ 
\frac{\lambda_3^2}{v_s} \right).
\ee
\be
\label{4}
\frac{d  \lambda_4}{d \ln L}=(1-K_c)\lambda_4+ 
\frac{\lambda_1 \lambda_4}{4\pi v_s}-
\frac{\lambda_3 \lambda_4}{2\pi v_s}.
\ee
\be
\label{5}
\frac{d  \lambda_5}{d \ln L}=(1-K_c) \lambda_5
-\frac{3}{16\pi v_s}\,\lambda_3\,
\lambda_4.
\ee
We  note the similarity of the above R.G. equations with those of 1D Kondo lattice
problem \cite{fujimoto}.
The Eq.(\ref{K},\ref{K2}) suggests a combination of fields 
$\phi_{\rho \pm}=\frac{1}{\sqrt{2}} ( \phi_{\rho 1} \pm \phi_{\rho 2})$. In terms of 
$\phi_{\rho \pm}$, the sum of the charge parts of our action becomes
\ba
\label{boson2}
S_c&=&\frac{ v_c}{ 2}(\frac{1}{ K_c}+\frac{1}{K_2})
 \int d x d \tau\,\Big[
\left(\frac{\partial \phi_{\rho +}}{\partial v_c \tau} \right)^2+
\left(\frac{\partial \phi_{\rho +}}{\partial x} \right)^2 \Big] \nonumber \\
&+&
\frac{ v_c}{ 2}\,(\frac{1}{ K_c}-\frac{1}{K_2}) \int d x d \tau \Big[
\left(\frac{\partial \phi_{\rho -}}{\partial v_c \tau}
 \right)^2+
\left(\frac{\partial \phi_{\rho -}}{\partial x}
 \right)^2 \Big].
\ea
Then from Eq.(\ref{K},\ref{K2}),
\be
\label{newK}
\frac{\partial (\frac{1}{ K_c}+\frac{1}{K_2})}{\partial \ln L}=0,\;\;
\frac{\partial (\frac{1}{ K_c}-\frac{1}{K_2})}{\partial \ln L}=
\frac{2 \lambda_4^2}{\pi v_c v_s}\,f_1(\frac{v_c}{v_s}).
\ee
We investigate the properties of the above R.G. equations in the next section.
\section{R.G. flow and phase diagram}
We consider only the ferromagnetic Hund's coupling constants,
 so that the couplings
$\lambda_i, i=1,\ldots,5$ should be taken to be positive 
in the physically relevant
range. 
Given the initial values of $K_c, \lambda_i, i=1,\ldots,4$, the R.G. flows
are uniquely determined. The initial values are determined by 
Fermi momentum $k_F$, $ U/t $, and $ J_H/t $. 
The derived R.G equations are valid  until 
$\max[\lambda_i/v_s] \sim O(1)$.  If all $\lambda_i$ converge to finite values
as $t=\ln L \to \infty$, the initial fixed point is stable and the R.G. equations
are valid along the whole R.G. trajectory 
since the $\lambda_i(\infty)$ can be made arbitrarily small by taking
sufficiently small $J_H$ and $U$ \cite{balents}. If any coupling constant diverges, 
the {\em asymptotic} behaviour of all coupling constants can be determined by R.G equations 
\cite{balents}.

Some properties of the R.G. equations can be understood by simple inspection
without numerical integration.
We find that $\lambda_2$ is marginal  from Eq.(\ref{2}). 
The marginality of $\lambda_2$ is
due to the chirality of interaction: it couples only the currents with
the same chirality.
The symmetric charge mode $\phi_{\rho +}$ is not  renormalized 
as can be seen in Eq.(\ref{newK}).  This 
is physically obvious since only the relative charge fluctuations affect
Hund's coupling. On the contrary, we see from Eq.(\ref{newK}) that the
Luttinger parameter of the asymmetric charge mode $\phi_{\rho
-}$ renormalizes to zero when the R.G. equations are 
extended beyond its validity range. This implies that $\phi_{\rho -}$ is pinned, 
and it is consistent with the formation of a  charge gap from $\lambda_4$ renormalization (see
below).
In Eq.(\ref{4}), 
the first term in the right hand side 
is the most dominant for the range $ J_H \ll U < t$, in which case 
$\lambda_4$ is certainly relevant. 
It gives  rise to {\em both} the charge gap of $\phi_{\rho -}$ and 
the {\em spin}
gaps of $\theta_{\sigma -}$ and $\phi_{\sigma +}$, as can be seen by 
expressing $\lambda_4$ term in the effective action in Abelian bosonized form
using Eq.(\ref{vertex}).
However, for the opposite range $ U \ll J_H  < t$, 
the first and the third terms in the right hand side compete, and
only the numerical integrations can determine the limiting behaviour. 
In this range,  $\lambda_4$ is expected to increase  slower compared with that
of the range  $ J_H \ll U < t$.
$\lambda_5$ starts to grow slowly initially since the initial value 
vanishes. But eventually the first term of Eq.(\ref{5}) dominates and $\lambda_5$
becomes relevant. 

The values of gaps can be estimated in the range 
$J_H \ll U < t$ by integrating the Eq.(\ref{4}) until the 
$\lambda_4$ grows up to the cut-off value $ t a$. 
\be
\label{gap}
\Delta_{c,s} \sim t  \Big[\frac{J_H}{t}\Big]^{\frac{1}{1-K_c}}.
\ee
Note that this estimated value of the gaps is larger for the stronger Hubbard repulsion.
For the range $U \ll J_H < t$, no analytical expression is available.
It is worth noting that the crucial  $\lambda_4$ interaction is not of the 
current-current interaction type, thus it is not constrained by the current 
conservation law and  allows the presence of the anomalous scaling dimension. 

We have numerically integrated R.G. equations for a range of different initial
values $U/t, J_H/t$.  
Some typical R.G. flows 
at $U=0$ and $k_F=\pi/4$ are shown in Fig.1 and Fig.2. 
The Fig.1 clearly shows the diverging $\lambda_4$ as $t$ increases for a range of
initial values of $J_H$. After the initial transient period 
all flows merge, which implies that the first term in Eq.(\ref{4}) dominates.
As mentioned above, for non-zero $U$, the divergence of
$\lambda_4$ is much faster. In Fig.2, the R.G flow for $U=0$ and a very small initial value of
$J_H=0.01$ is shown. Initially, the $\lambda_4$ slightly decreases due to the 
last term in R.G Eq.(\ref{4}), but soon it begins to increase due to the 
first term of Eq.(\ref{4}), and eventually diverges. From the examination of
R.G flows, we conclude that the charge and spin gaps open up for {\em arbitrarily} small
initial values of $J_H$ irrespective of $U$. 
Our system Eq.(\ref{H1}) is in the massless phase only for a line $J_H=0$, 
which is simply two decoupled Hubbard chain, 
where each of Hubbard chain is in critical Luttinger 
liquid phase.
In any other region with positive $J_H>0$, $\lambda_4$ is  relevant and the charge
and spin gaps are present.

\section{Discussions}
The spin sector of a single Hubbard  chain at low energy is in the same universality class
of antiferromagnetic(AF) XXX Heisenberg chain \cite{boso1}. 
Therefore, the spin sector of the system 
Eq.(\ref{H1}) can be approximately described as two spin 1/2 XXX chains coupled
by ferromagnetic exchange coupling at low energy. 
In the  strong coupling limit $J_H \to \infty$ we only need to
consider the triplet combination of spins from each spin chain. When projected 
onto this triplet subspace, the coupled spin 1/2 chains become essentially single
spin 1 chain. The spin 1 chain is well known to be gapped \cite{boso1},
 and the gap is called
Haldane gap \cite{haldane}.
Also Strong and Millis \cite{strong} studied the coupled spin 1/2 chains 
and showed that  the sping gap opens even for infinitesimally small 
$|J_H|$, which is consistent with our result.  
The spin gap we have obtained can be also understood from the semiclassical point
of view \cite{nagaosa}. 

It is interesting to examine other parameter domains 
even though they are not directly related to our system.
First, the antiferromagnetic exchange can be considered. It corresponds to a 
negative $J_H$, which implies the signs of $\lambda_i, i=2,3,4,5$ to be flipped.
The R.G equation of charge Luttinger parameters, Eq.(\ref{newK}), is unchanged,
but $\lambda_3,\lambda_4,\lambda_5$ now become relevant. 
$\lambda_3$ only generates  a spin gap, while $\lambda_4, \lambda_5$ generate
both the charge and spin gap. 
The relative initial values of the couplings determine the gap which is going 
to be opened the fastest, and R.G flows depend critically on the gap. 
The detailed investigation will not be presented here.
Second, the attractive Hubbard interaction is another possibility (with positive
$J_H$). In this case,  the sign of $\lambda_1$ is flipped and the initial value
of $K_c$ is {\em greater } than 1.  In this case, $\lambda_1$ becomes marginally 
relevant, and  $\lambda_4$ becomes irrelevant.
The marginally relevant $\lambda_1$ would open up a {\em spin } gap for
{\em each} band $l=1,2$. 
This is essentially the Luther-Emery spin gap \cite{luther}, and
the Hund's coupling does not play any role here.

In this paper we have investigated two
Hubbard chains coupled by Hund's coupling  using renormalization
group method for arbitrary densities away from half filling. 
We find that in the weak coupling regime
Hund's coupling is always relevant, irrespective of the stength of $U$, 
and opens up gaps for both symmetric and antisymmetric spin modes
and for antisymmetric  charge mode.

\bigskip
\centerline{{\bf ACKNOWLEDGEMENTS}}
H.C. Lee is grateful to Prof. A. Changrim for discussions.
This work was  supported by 
Korea science and engineering foundation through the 
Quantum-functional Semiconductor Research Center at Dongguk University.


\begin{figure}
\caption{R.G. flow for three different values of $J_H=0.4, 0.2, 0.05$ (Here $U=0$ and
 $J_H$ is measured in units of $t$). }
\end{figure}

\begin{figure}
\caption{R.G. flow for  $J_H= 0.01$ (Here $U=0$ and
 $J_H$ is measured in units of $t$). }
\end{figure}

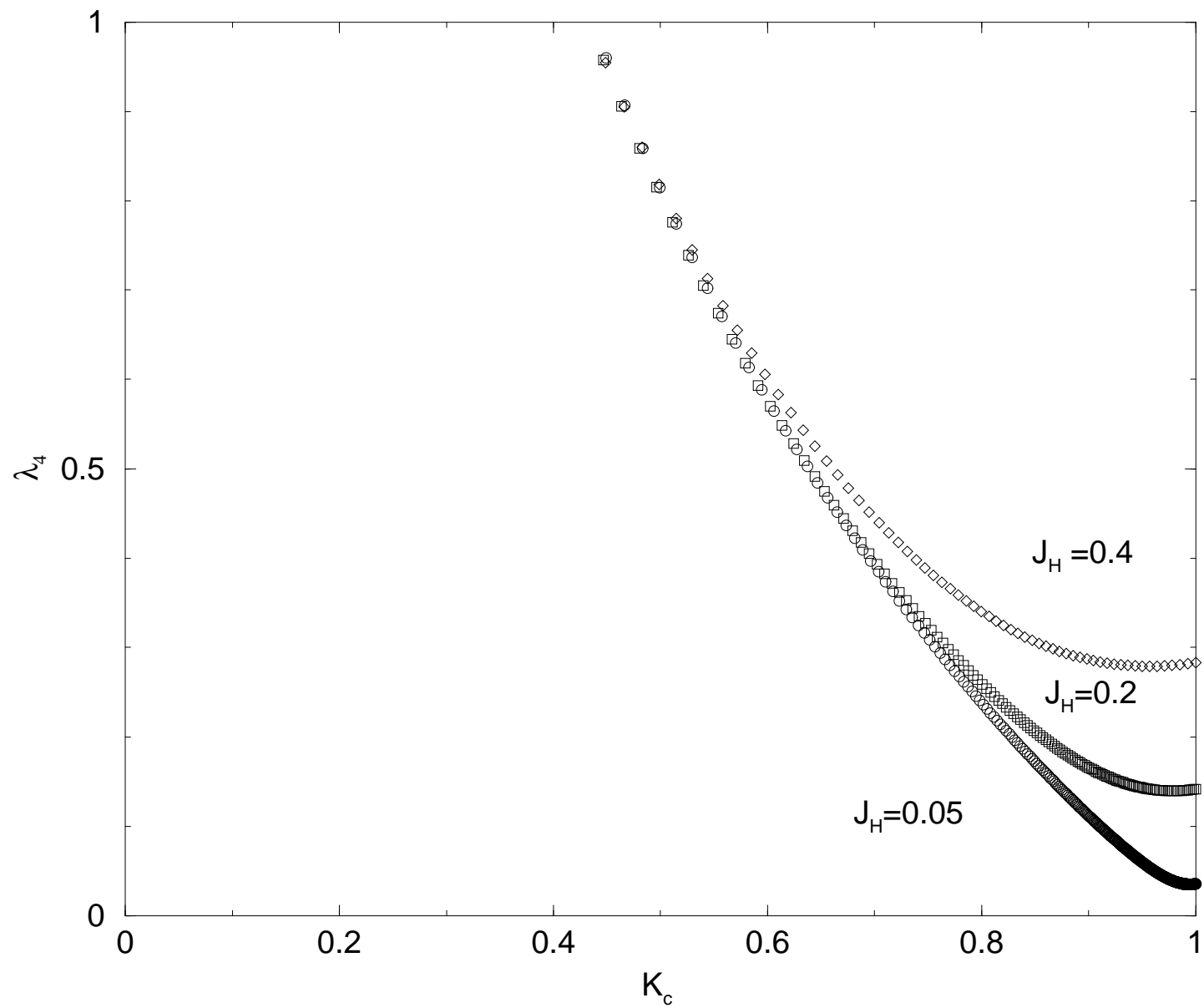
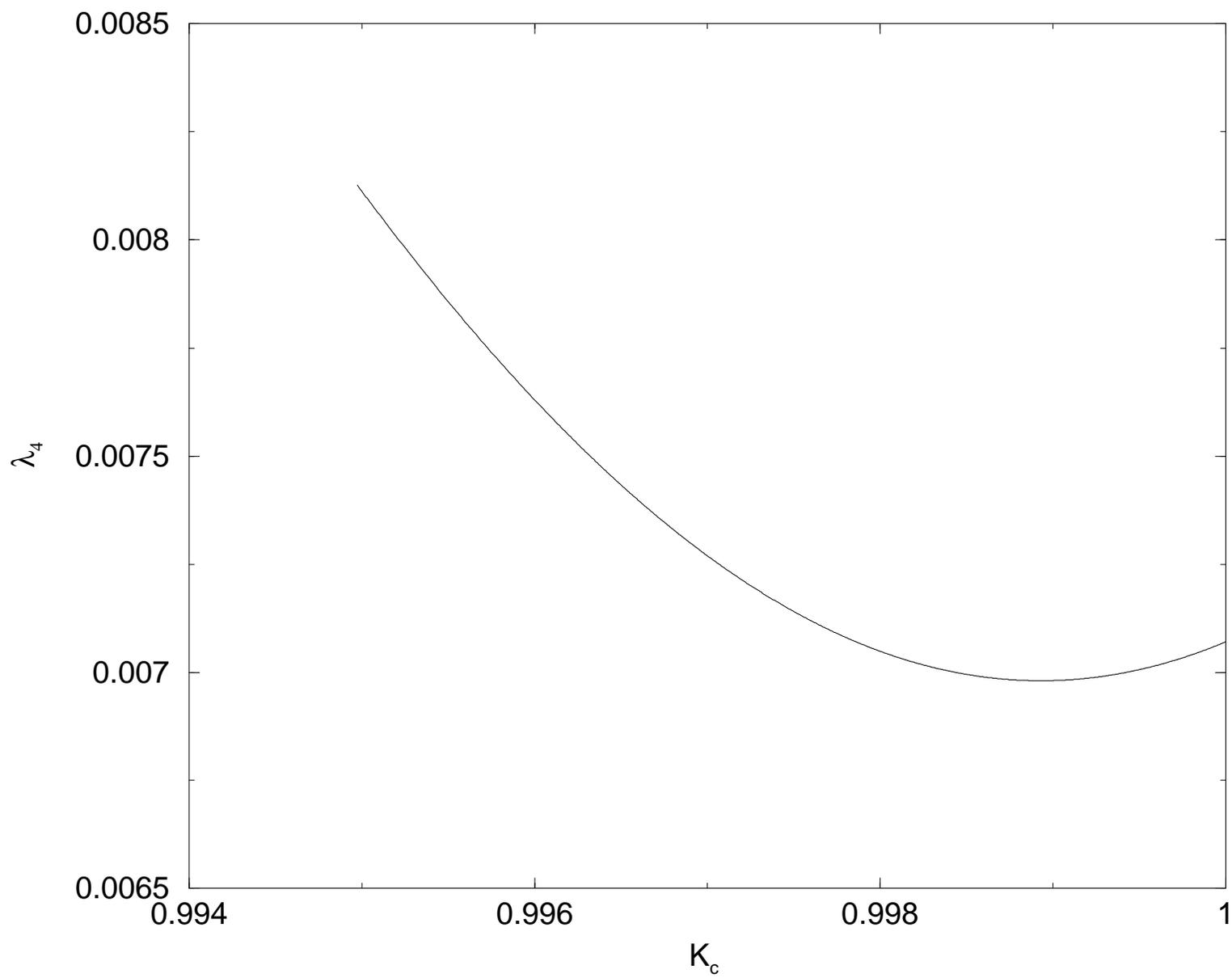
\end{document}